\documentclass[twocolumn, prd, nofootinbib, showpacs, amsmath, amssymb] {revtex4}

\begin{document}

\title{Do we have much freedom in defining of elementary KK spectrum?}

\author{Michael~Maziashvili}
\email{maziashvili@iphac.ge}\affiliation{ Andronikashvili
Institute of Physics, 6 Tamarashvili St., Tbilisi 0177, Georgia }

\begin{abstract}

Recently the spectrum of KK modes in the framework of one flat
extra-dimensional scenario was revisited in the paper Phys. Rev.
D74 (2006) 124013, (hep-th/0607246) on the basis of self-adjoint
extension of the quantum mechanical operator determining the KK
masses. In this Letter we note that the range of allowed boundary
conditions on the KK modes is overestimated in above mentioned
paper and give all allowed possibilities.

\end{abstract}



\widetext

\maketitle

In ADD scenario \cite{ADD} the fundamental scale of gravity is
taken to be close to the electroweak scale, $M_F \sim$TeV, in
presence of compact extra dimensions. Denoting the size of extra
dimensions by $R$ one finds that the volume of extra space $R^n$,
where $n$ denotes the number of extra (spatial) dimensions,
determines the fundamental scale through the Planck one

\[  R^n M_F^{n+2} \sim M_P^2~,
\] which in the case $M_F \sim $TeV yields

\begin{equation}\label{size} R \sim 10^{30/n - 17} \mbox{cm}~. \end{equation}
In the framework of simple compactifaction scheme used in
\cite{ADD} the mass gap between zero and first KK modes is given
by $R^{-1}$ and correspondingly the size $R$ determines the scale
at which the Newtonian inverse square law (valid for $r \gg R$)
starts to change for the higher dimensional expression. In view of
this relation, for $n=1$ from Eq.(\ref{size}) one finds the size
of extra dimension to be unacceptably large (as to modify
Newtonian gravitation at solar-system distances) $R \sim
10^{13}$cm. For $n=2$, however, the extra dimension is at a
millimeter scale $R \sim 1$mm, order of magnitude greater than the
scale of present experimental bound for the Newtonian inverse
square law $\sim 0.1$mm.

 An interesting point in regard with the ADD model was stressed by Dines
\cite{Din} emphasizing the role of shape moduli of the compact extra space.
It is important point as the shape moduli also play a significant
role in determining the experimental bounds on such scenarios.
Keeping the fundamental scale in the TeV range by fixing the
volume of extra space there still remains a freedom due to
geometry of extra space which allows one to increase the mass gap
of KK spectrum arbitrarily. This observation alleviates many of
the experimental bounds on ADD model, that is, it makes extra
dimensions essentially invisible with respect to experimental
constraints on the light KK modes. This approach works for $n \geq
2$ since one-dimensional compactifications lack shape moduli.

 It was noticed in paper \cite{GN} that there still remains a freedom in
defining of KK spectrum due to self-adjoint extension of quantum mechanical
operator determining the masses of KK modes. It would be interesting if at the
expense of self-adjoint extension parameter(s) one could retain the zero mode
but increase the mass gap between zero and first KK modes in such a way to
alleviate some of the experimental bounds or to avoid for instance the modification of gravity in ADD scenario with one extra dimension at the
solar-system scale. Guided by this idea let us look how much freedom do we
have in this way. To quantify
the discussion let us consider a neutral scalar field in presence of one flat spatial extra
dimension \begin{equation}\label{higherac} S[\varphi] = \int\limits_0\limits^Rdz \int d^4x\,\partial_A\varphi\partial^A\varphi~.\end{equation} The standard way is to decompose a
higher dimensional field into a complete set of functions
$\varphi = \sum\limits_n \phi_n(x)\xi_n(z)$ where \begin{equation}\label{massop}-\partial^2_z\xi_n(z) =
m^2_n\xi_n(z)~,\end{equation} and make a reduction of the action
(\ref{higherac}) with respect to this decomposition \cite{Rizzo}. With no loss of generality one can assume the functions
$\xi_n(z)$ to be real and satisfy the condition \[\int\limits_0\limits^Rdz\,
\xi_m(z)\xi_n(z) = \delta_{mn}~. \] Substituting this decomposition into Lagrangian from Eq.(\ref{higherac}) one
gets
\begin{eqnarray}\label{KKaction}S[\varphi] = \sum\limits_n\int d^4x
\,\left\{\partial_{\mu}\phi_n\partial^{\mu}\phi_n -m^2_n\phi^2_n\right\} \cr
-\sum\limits_{m,n}\,( \xi_n\partial_z\xi_m)\Big|^R_0\times  \int
d^4x\, \phi_n\phi_m ~. \end{eqnarray} To get rid of the last term in
Eq.(\ref{KKaction}) one has to require
\begin{equation}\label{boundterm} \left(\xi_n\partial_z\xi_m +
      \xi_m\partial_z\xi_n \right) \Big|^R_0  = 0~. \end{equation} Under this
  boundary condition the action (\ref{KKaction}) is reduced to the sum of
  independent 4D scalar fields $\phi_n$ with masses $m_n$, a so called KK tower. Certainly the boundary
condition (\ref{boundterm}) does not specify uniquely the wave function $\xi_n$
and corresponding mass spectrum. Therefore one needs some physical principle
that will guide into this problem. Due to standard quantum
mechanical ideology the physical quantities are represented by the
linear, self-adjoint operators in the Hilbert space. In what follows we
require the operator $-\partial^2_z$ determining the KK mass spectrum through
the Eq.(\ref{massop}) to be self-adjoint \cite{RS}. In general a self-adjoint extension
consists in defining a maximal domain of the Hermitian operator coinciding
with the domain of its Hermitian conjugate. Hermiticity of the operator
$-\partial^2_z$, that is, \[\int\limits_0\limits^Rdz \, \xi_n\partial^2_z\xi_m
= \int\limits_0\limits^Rdz \, \xi_m\partial^2_z\xi_n~,\] implies
\begin{equation}\label{hermcon} \left(\xi_n\partial_z\xi_m -
    \xi_m\partial_z\xi_n \right) \Big|^R_0  = 0~. \end{equation} Combining
the conditions (\ref{boundterm}) and (\ref{hermcon}) one gets \begin{equation}\label{finboundcond} \xi_m\partial_z\xi_n \Big|^R_0  = 0~. \end{equation}
This condition immediately implies \[m_n = - \int\limits_0\limits^Rdz
\,\xi_n\partial^2_z\xi_n = \int\limits_0\limits^Rdz
\, \left(\partial_z\xi_n\right)^2 -  \xi_n\partial_z\xi_n \Big|^R_0 \geq
0~. \] It is easy to see that the following boundary conditions
\begin{eqnarray}\xi_n(0) & =& \pm \xi_n(R)~,~~\partial_z\xi_n(0) = \pm
  \partial_z\xi_n(R)~;\nonumber \\ \xi_n(R)& =& 0~,~~\partial_z\xi_n(0) = 0~;
  \nonumber\\ \xi_n(0)& =& 0~,~~\partial_z\xi_n(R) = 0~;
  \nonumber\\
 \xi_n(0)& =& \xi_n(R) = 0~;\nonumber \\ \partial_z\xi_n(0)& =& \partial_z\xi_n(R) =
 0~.\nonumber \end{eqnarray} are appropriate with Eq.(\ref{finboundcond}). One can write
the Eq.(\ref{finboundcond}) in the following general form
\begin{equation}\label{oneparfamily}\xi_n(0) = \alpha \xi_n(R)~,~~ \partial_z\xi_n(R) = \alpha
  \partial_z\xi_n(0)~,  \end{equation} where $\alpha$ is a real
parameter. The Eq.(\ref{oneparfamily}) does not comprises any boundary
condition containing either $\xi_n(R) = 0$ or $\partial_z\xi_n(0) = 0$. The
solution of Eq.(\ref{massop}) with nonzero $m_n$ reads \[\xi_n = a_n\sin(m_nz) +
b_n\cos(m_nz)~,\] while for the zero mode, $m_n = 0$, it takes the form \[\xi_0
= a\,z + b~.\] It is easy to see that the presence of zero mode requires
$\alpha = 1$ and correspondingly at the expense of self-adjoint extension one
can not increase a mass gap between zero and first KK modes. It is easy to show that the boundary conditions
(\ref{oneparfamily}) imply \[\cos(m_nR) = {2\alpha \over 1+ \alpha^2}~,\]
which for a given value of $\alpha$ leads to the uniform uplifting of the KK
masses considered in ADD scenario. The boundary conditions that don't belong to
the one-parameter family (\ref{oneparfamily}) can be easily analyzed:
\[\xi_n(R) = \partial_z\xi_n(0) = 0~,~ \Rightarrow ~m_n={1\over R}\left({\pi\over
  2}+ \pi n\right)~,~n \geq 0~, \]  \[  \xi_n(0) = \xi_n(R) = 0~,~\Rightarrow ~
m_n = {\pi n\over R}~,~n \geq 1~,\]\[ \partial_z\xi_n(0) = \partial_z\xi_n(R) =  0~,~\Rightarrow~m_n = {\pi n\over R}~,~n \geq 0~.\]

For the sake of safety one may prefer to write first the variational problem
for higher-dimensional theory and then use the decomposition of the field. The variation of action (\ref{higherac})
gives

\begin{equation}\label{variation} \left\{
\begin{array}{ll}  \partial_A\partial^A\varphi = 0~,
\\ \int\limits_0\limits^R dz \int d^4x\,\partial_A
\left(\delta\varphi\partial^A\varphi \right) = 0 ~. \end{array}\right.\end{equation}
Using the decomposition \[\varphi =
\sum\limits_n \phi_n(x)\xi_n(z),~~~~~\delta\varphi =
\sum\limits_n \delta\phi_n(x)\xi_n(z)~,\] for the boundary term (second
equation in (\ref{variation})) one finds

\begin{eqnarray} \label{varboundterm} \sum\limits_{n,m}\left[\int\limits_0\limits^{R}dz\, \xi_n\xi_m
    \int d^4x\, \partial_{\mu}\left(\delta \phi_n\partial^{\mu}\phi_m\right)
 \right. ~~~~~~~~~~ \nonumber \\ \left. - \int d^4x\, \delta\phi_n
   \phi_m\int\limits_0\limits^{R}dz\,\partial_z\left(
     \xi_n\partial_z\xi_m\right) \right] = 0~. \end{eqnarray} Imposing a
standard boundary condition: $\delta\phi_n = 0$ at the infinity of
the $x^{\mu}$ space, one finds that the first term in Eq.(\ref{varboundterm})
vanishes and what remains due to arbitrariness of $\delta\phi_n$ is just the
condition (\ref{finboundcond}), which satisfies the condition
(\ref{hermcon}) and therefore immediately guarantees Hermiticity of the operator $-\partial_z^2$.

To summarize, we have considered two approaches to the KK problem
in a simplest ADD compactification scheme in order to see whether
a self-adjoint extension of the quantum mechanical operator
determining KK mass spectrum can lead to some new physical
insights. Firs we decomposed the higher dimensional field into the
complete set of KK modes and by substituting this decomposition
into the action studied the variational problem. Then requiring
simply the Hermiticity of a quantum mechanical operator
determining KK mass spectrum we got the general boundary condition
(\ref{hermcon}) for KK modes. Let us notice that in their study
the authors of \cite{GN} have not taken into account the condition
(\ref{boundterm}) and correspondingly overestimated the range of
allowed boundary conditions on $\xi_n$ functions. That is, instead
of the boundary condition (\ref{finboundcond}) they are dealing
with the Eq.(\ref{hermcon}) which may contain $\xi_n$ functions
that simply do not satisfy the variational problem.

In our second approach we started directly from the
higher-dimensional variational problem resulting correspondingly
in equation(s) of motion + boundary conditions. To satisfy this
variational problem in the form of KK decomposition one
immediately arrives at the same boundary condition
(\ref{finboundcond}) that thus determines all freedom in defining
KK spectrum. Analyzing this boundary condition, we arrived at a
conclusion that in the ADD compactification scheme at the expense
of a self-adjoint extension of quantum mechanical operator
determining the KK spectrum one can not increase a mass gap
between the KK modes that could lead to some new physics. Thus we
see the misleading results of \cite{GN} are due to ignorance of
variational problem (that is usually implied by the action) that
besides the equations of motion implies the boundary conditions
that should be also satisfied.

\vspace{0.5cm}

\begin{acknowledgments}

It is a pleasure to thank Professors Jean-Marie Fr\`{e}re and Peter Tinyakov
for invitation and hospitality at the\emph{ Service de Physique Th\'eorique,
  Universit\'e Libre de Bruxelles}, where this Letter was written. Thanks are
due to Peter Tinyakov and Michel Tytgat for useful discussions and comments. The work
was supported by the \emph{INTAS Fellowship for Young Scientists} and
the \emph{Georgian President Fellowship for Young Scientists}.

\end{acknowledgments}

\end{document}